# Investigating Analyte Co-Localization at Electromagnetic Gap Hot-Spots For Highly Sensitive (Bio)molecular Detection by Plasmon Enhanced Spectroscopies

*Rishabh Rastogi,[†,‡] Hamed Arianfard,[§] Prof. D. Moss,[§] Prof. S. Juodkazis,[§] Prof. P. Michel-Adam,[‡] Dr. S. Krishnamoorthy[†]\**

[†] Materials Research and Technology (MRT) Department, Luxembourg Institute of Technology, 41, Rue du Brill, Belvaux, L-4422 Luxembourg, [‡] Laboratory Light, Nanomaterials & Nanotechnologies – L2n, the University of Technology of Troyes & CNRS ERL 7004, 12 rue Marie Curie, 10000 Troyes, France, [§] Optical Sciences Centre, Swinburne University of Technology, Hawthorn, VIC 3122, Australia

Electromagnetic hot-spots at ultra-narrow plasmonic nanogaps carry immense potential to drive detection limits down to few molecules in sensors based on surface enhanced Raman or Fluorescence spectroscopies. However, leveraging the EM hot-spots requires access to the gaps, which in turn depends on the size of the analyte in relation to gap distances. Herein we leverage a well-calibrated process based on self-assembly of block copolymer colloids on full-wafer level to produce high density plasmonic nanopillar arrays exhibiting large number (> $10^{10}$ cm$^{-2}$) of uniform inter-pillar EM hot-spots. The approach allows convenient handles to systematically vary the inter-pillar gap distances down to sub-10 nm regime. The results show compelling trends of the impact of analyte dimensions in relation to the gap distances towards their leverage over inter-pillar hot-spots, and the resulting sensitivity in SERS based molecular assays. Comparing the detection of labelled proteins in surface-enhanced Raman and metal-enhanced Fluorescence configurations further reveal the relative advantage of Fluorescence over Raman detection while encountering the spatial limitations imposed by the gaps. Quantitative assays with limits of detection down to picomolar concentrations is realized for both the small organic molecules and the proteins. The well-defined geometries delivered by nanofabrication approach is critical to arriving at realistic geometric models to establish meaningful correlation between structure, optical properties and sensitivity of nanopillar arrays in plasmonic assays. The findings emphasize the need for the rational design of EM hot-spots that take into account the analyte dimensions to drive ultra-high sensitivity in plasmon-enhanced spectroscopies.

Optical sensors that employ plasmonic transducers have gained significant attention in the last decade, especially due to their ability to detect analytes at ultra-low concentrations, in miniaturized volumes, with application to a range of sectors including medical diagnostics,[1,2] environmental monitoring,[3,4] food safety,[5,6] defense,[7,8] and forensics.[9,10] Plasmon-enhanced spectroscopies sensors such as those relying on electromagnetic enhancement to Raman or Fluorescence intensities of analyte were shown to be particularly promising in driving analytical detection down to single-molecule level.[11–14] However, despite over four decades of work in the direction, their adoption into practical technologies remains limited, especially due to the inability to assure high sensitivity without compromising reliability in performance and scalability in production.[15] Sensitivity in plasmon-enhanced spectroscopy results from a disproportionately high enhancement of electromagnetic (EM) field at the vicinity of optically excited metal nanostructures, e.g. metal nanogaps, sharp metal ends with high curvatures (lightning rod effect).[16–18] Fang et al showed that only 63 molecules per million were required to achieve a sizeable signal (24% of signal contribution) when present within EM hot-spots presenting enhancement factors >$10^9$.[19] Such high EM enhancements are achievable at ultra-narrow gaps or junctions between adjacent metal nanostructures typically for gap distances approaching sub-10nm regime.[20–23] While such gaps can accommodate analytes smaller than the gap size, they would be incapable of accommodating large molecules, e.g. proteins, or oligonucleotides. The ensuing spatial constraints towards leveraging the highly enhanced EM hot-spots for biosensing negatively impacts the sensitivity of plasmonic bioassays. While there has been intensive research in the past decade on approaches to achieve highly enhancing EM hot-spots, their performance has often been validated using small molecules, typically with sub-nanometric dimensions. While such validation helps to understand the enhancement at the EM hot-spots, they do not address the challenge of the inability to accommodate larger biomolecular analytes.

Different approaches were adopted in literature to achieve analyte co-localization at EM hot-spots. In one of the studies, the irreversible collapse of high-aspect-ratio metal nanopillars was used to mechanically trap molecular analytes at the inter-pillar EM hot-spots.[24] However, the mechanical flexibility of the pillars also risks adversely impacting their stability, as they cannot be subjected to multiple washing-drying steps, requires storage in scrupulously moisture-free conditions, and cannot be subjected to classical multi-step adsorption sequence typical of most biological assays. Further, the approach is devoid of opportunities for the rational design of the geometry of the EM hot-spots. Another approach includes the formation of vertical 3D hot-spots defined by the analyte sandwiched between a metal nanoparticle and a plasmonic substrate.[25] The width of the vertical EM hot-spots is defined by the size of the sandwiched analyte, and thus is sensitive only for small molecules but not large biomolecules whose detection relies on receptor-ligand complexes with dimensions spanning several tens of nanometers.[26] In an instance that resembles this strategy, plasmonic sensing of a cancer biomarker was demonstrated by sandwiching an antibody-antigen complex (~30-40 nm) between detection antibody labeled with gold nanostars and gold nanotriangle array.[27] The approach, however, gains largely from a high concentration of highly sensitive dye reporters trapped within the gold nanostars, rather than the EM enhancements between the nanostars and the metal nanoarray. Alternative configurations have been shown using dynamic plasmonic nanostructures that can swell and collapse in response to the presence of an analyte. These configurations have been typically demonstrated for the detection of physiological changes[28–31] such as temperature and pH, or small molecules such as water (moisture), or glucose, but not been exploited for sensing larger molecules.[29,32] These evidences reinforce the challenge of configuring EM hot-spots that take into account the large dimension of biomolecules. The advancement in this direction requires the ability to perform reliable investigations of the sensitivity of plasmonic transducer as a function of the geometry of the EM hot-spot while taking into account the molecular dimensions. This will however require producing well-defined EM hot-spots with geometries precisely controlled down to sub-molecular dimensions to serve as good models to derive robust correlations. Further, the EM hot-spots should be obtained on large enough numbers to allow systematic investigations, and over macroscopic areas that may permit use of alternative characterization tools with a macroscopic footprints (e.g. XPS, Ellipsometry) if so desired. Herein we show a clear impact of the molecular dimension of analytes in relation to that of EM hot-spots in the ability to leverage EM enhancements at inter-pillar hot-spots for SERS and MEF assays. The choice of geometry and assay configuration is shown to deliver analytical sensitivities down to picomolar concentration for analyte smaller and larger than the EM hot-spots.

**EXPERIMENTAL SECTION**

*Materials:* Polystyrene-block-poly(2-vinylpyridine) (PS-*b*-PVP) (102000-b-95000 $g$ $mol^{-1}$) was purchased from Polymer Source Inc. (Montreal, Canada). Silicon (Si) substrates with thick Silicon Oxide ($SiO_2$) coatings were purchased from Silicon Valley Microelectronics (Santa Clara, CA). 1-Napthalenethiol (≥99%), *m*-Xylene (≥99%) were purchased from Sigma-Aldrich. Biotin-PEG-Thiol (BPT) (400 $g$ $mol^{-1}$) was purchased from Nanocs, New York, USA. Phosphate Buffer Saline (PBS) and Bovine Serum Albumin (BSA) solution (*10x*) was purchased from R&D Systems, Abingdon, UK. Streptavidin Conjugated with Cyanine-5 dye was purchased from Rockland, Limerick, PA. AFM Tips were purchased from Nanoandmore, Germany. Polydimethylsiloxane (PDMS) was purchased from Dow Corning along with the activator to form PDMS blocks.

*Nanopillar Arrays:* The pillar arrays were prepared using a protocol that was reported earlier. The coating solution was prepared using PS-*b*-PVP with a molecular weight of 197 kDa in a *p*-Xylene organic solvent (*0.8% w w⁻¹*). These organic templates were spin-coated at 6000 rpm over freshly cleaned 100 mm Si wafer consisting of 25 nm $SiO_2$ film. $C_4F_8/CH_4$ gas plasma to transfer the template pattern into the underlying $SiO_2$ layer. Resulting $SiO_2$ pattern behaves as a highly selective pattern for Si etching using $SF_6/C_4F_8$ plasma to obtain silicon nanopillars. The center-to-center periodicity of 105.8 nm (coefficient of variation below 10% from chips to chips) of the features was preserved from the polymeric template, confirmed using AFM and SEM images from the template and nanopillar arrays. Edge separations between the gold-coated pillars was determined by subjecting the SEM images of the Au-NPA to image analysis using ImageJ.[47] The silicon nanopillars were sputter-coated while rotating at 10 rpm continuously with a 5 nm ±1 nm adhesive layer of Chromium at 100 W (60 sccm) of Argon (Ar) Gas under $10^{-2}$ mbar pressure for 14 s (0.36 nm s$^{-1}$). The desired thickness of Gold was deposited under similar conditions at 0.86 nm s$^{-1}$ rate with a standard deviation of 6%.

*Optical and Spectroscopic Measurement:* The absolute reflection measurements were taken from UV/Visible spectrometer (Perkin Elmer L1050) in an integrated sphere configuration using reflectance diffusive Teflon reference standard (WS-1, Ocean Optics). The resulting reflectance was automatically corrected with respect to the reference standard. SERS/MEF measurements were performed using the inVia confocal Raman instrument (Renishaw, UK) fully automated with the excitation LASER of 633 nm and 50X/0.55 N.A. long-distance Leica objective lens. For NT Raman probe detection, an average of 10 scans per spot was acquired and an average of 10 different spots on each sample, to calculate the standard deviations. Each scan was taken with a total exposure of 10s with a 633 nm He-Ne LASER at a power of 1.9 mW. For bio-molecular detection using fluorescent-dye Cy5 conjugated to Streptavidin, each measurement was recorded at 5 different places throughout the functionalized surface, with 3 s of total exposure time at varying power (0.0095 to 0.095 mW) to prevent the saturation of the CCD due to Fluorescence background. Baseline subtractions were performed according to the default settings present in the Renishaw Raman software program Wire 5.0 and all the scans were later normalized with respect to power and exposure time to acquire the absolute signal for comparison between different settings in OriginPro software.

*Plasmonic Assays:* For NT assays, the Au-NPA were incubated overnight in the ethanolic solution of NT at a varying concentration from nM to µM range and were then rinsed thoroughly with absolute ethanol before drying and recording Raman spectra. Biotin-Streptavidin Interactions: Au-NPA was immersed in the 1 mM solution of Bi-functional PEG molecule (Biotin-PEG-Thiol, 0.4 kDa) for 2 hours. Then, the substrate was left for 30 minutes in BSA (1%) in PBS to block all the rest of the available sites on the gold-coated surface. These substrates are then covered with the freshly made PDMS block punched with "wells" of 5mm diameter. These "wells" were then filled with 30 μl of Streptavidin protein conjugated with Cy5 dye in varying concentrations and left for 2 hours. Later, these "wells" were rinsed properly with PBS and water before removing the PDMS block. PDMS block was prepared using the SYLGARD™ 184 Silicone Elastomer (Dow Corning). The wells allowed selectively functionalizing the regions on the same substrate maintaining low reagent consumption and ease of washing. The LOD was determined from the mean and standard deviation of the blank and the standard deviation of the signal at the lowest concentration. [48]

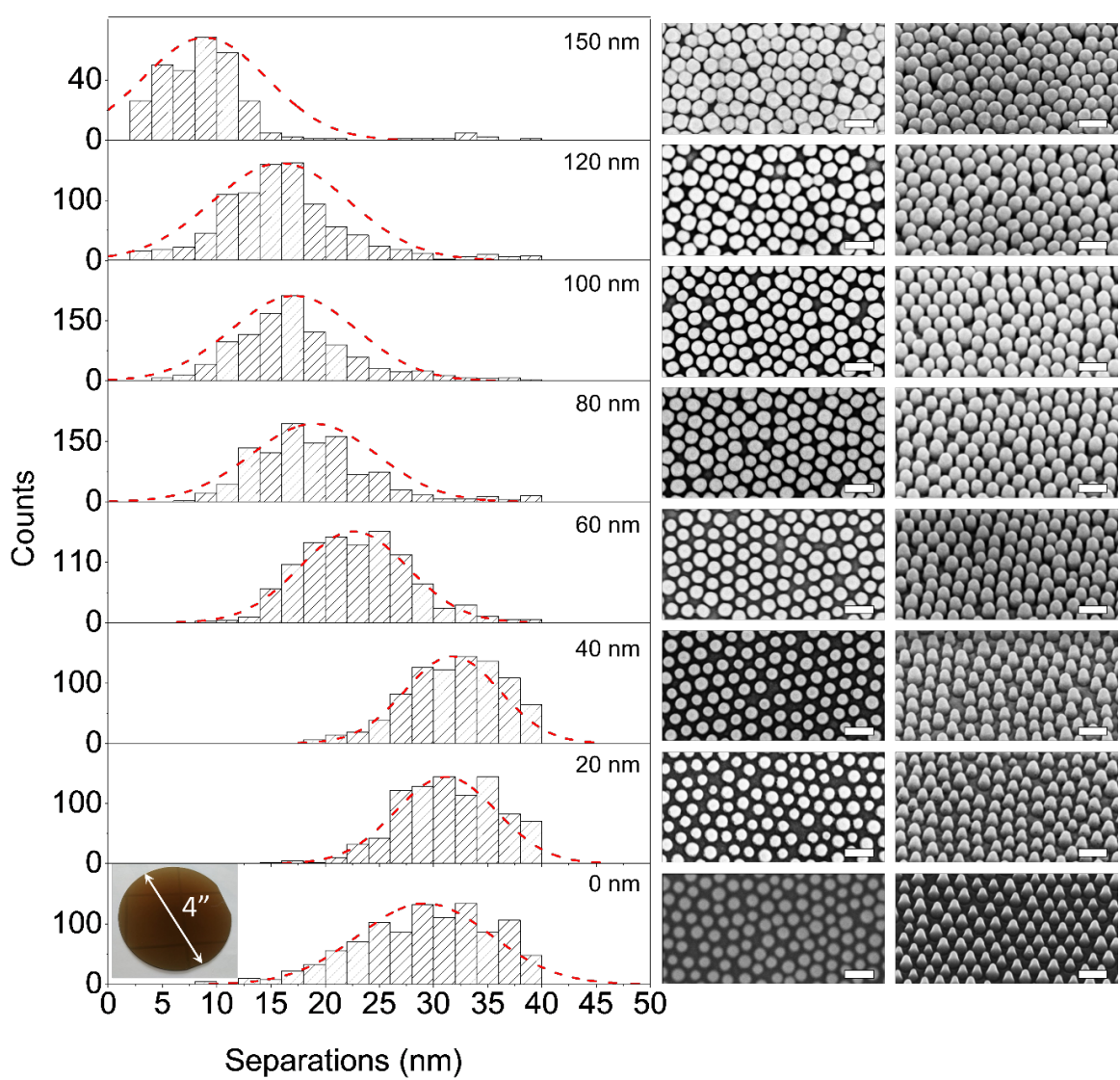

**Figure 1.** Histograms of inter-pillar gap distances as a function of the thickness of sputtered gold films, and the SEM images of the nanopillar arrays acquired in the top and tilt view (30°

inclination). The scale bars indicate 200 nm. Picture of 4" uniform nanopillar arrays on 4" Si wafers is shown as an inset.

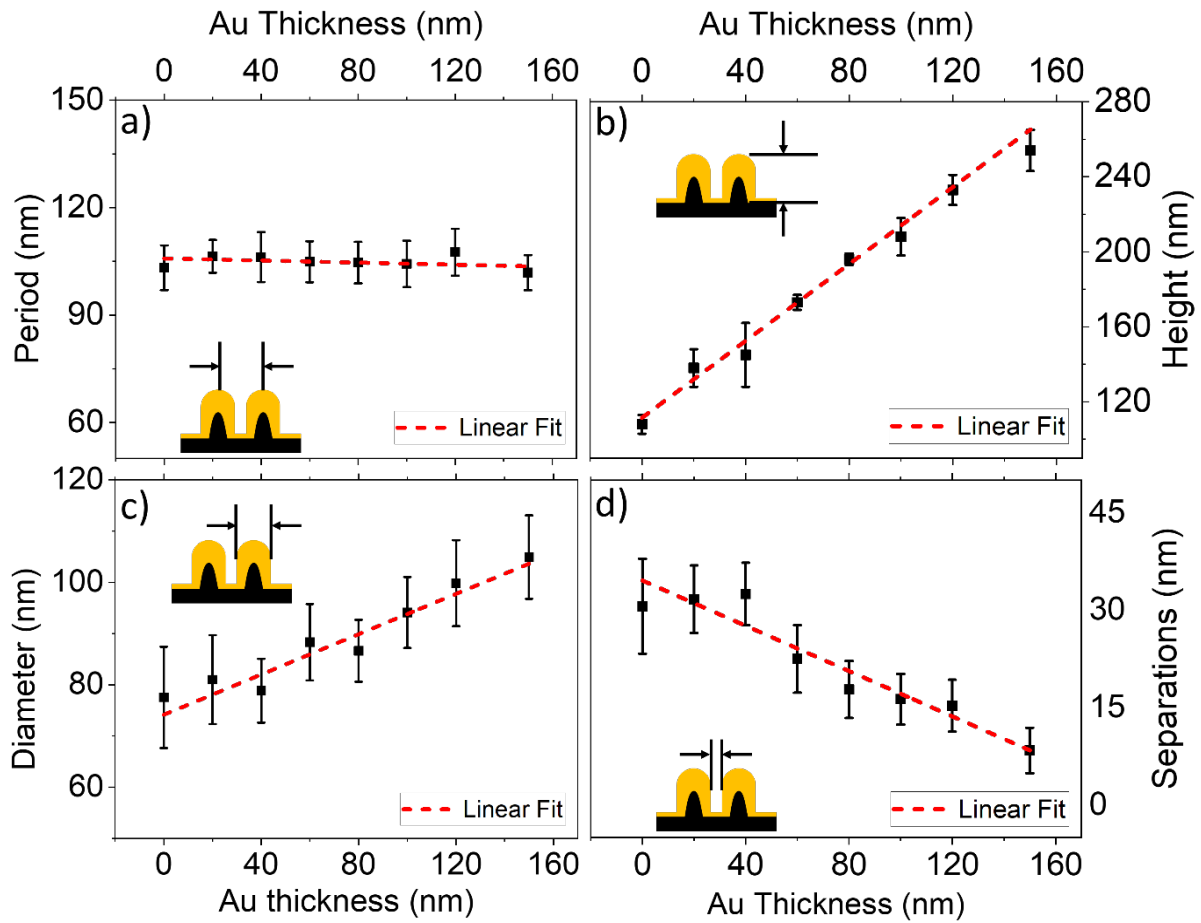

**Figure 2.** Plots show systematic variation in a) period, b) height, c) diameter and d) separations of Au-NPA as function of sputtered gold thickness.

**RESULTS AND DISCUSSION**

**Nanoengineered Inter-Pillar gaps.** The gold-coated nanopillar arrays exhibiting high density well resolved EM hot-spots were obtained by molecular self-assembly approach, reported earlier.[33,34] Ordered nanopillar arrays (NPA) were prepared by lithographic pattern-transfer of nanoscale polymer template obtained from soft colloidal particles of polystyrene-block-polyvinylpyridine into underlying silicon surface. The self-assembly approach gives control over template separations down to sub-10 nm length scales with uniformity spanning full wafers, and produced at high throughput.[33,35] The capability to achieve highly resolved geometries together with scalability across several square centimeters makes the approach unique and attractive in application to nanoplasmonic interfaces.

Nanopillars with a positively tapered profile with a height of 108 nm ± 5 nm, the diameter of 77.5 nm ± 9.8nm, and edge-to-edge separation of 30.6 nm ± 7.3 nm were obtained. 100 mm Silicon wafer with silicon nanopillars was diced into 1x1 cm chips for further use. The silicon NPA is subsequently sputter-coated with an adhesive layer of chromium followed by gold (Au) with thickness ranging from 20-150 nm to obtain ordered Au nanopillar arrays (Au-NPA). Cross-sectional measurements in SEM show conformal coating of gold on the silicon nanopillars in all cases resulting in systematic increase of the height and diameter, and decreasing inter-pillar separations in the Au-NPA. (Figure 1) An asymmetric growth of gold is observed, with the preferential increase in feature heights over diameters in steps of ~1.0 nm and ~0.2 nm respectively for every nanometer increase in thickness of sputtered gold. (Figure 2) The increase in the diameter of pillar consequently decrease the inter-pillar separations. At the early stages of sputter deposition, the diameter as viewed from the top-view SEM images do not change, and consequently yielding similar inter-pillar separations until thickness of 40 nm. This results from a transition of shape from the positively tapered silicon pillars into Au-NPA with a cylindrical side wall morphology. There is no apparent change in profile observed for thickness above 40 nm. The Au-NPA separation was found to systematically decrease in steps of ~ 0.2 nm for every nanometer of sputtered gold, resulting in values of separation in the range of 30.6 nm - 8.4 nm. (Figure 2) The histograms in Figure 1 show low standard deviation of the inter-pillar separations. Ability to realize well-defined Au_NPA with systematic changes in geometric parameters in sub-nanometric steps, is a crucial step forward while addressing geometric impact as function of analyte dimensions on sensitivity of plasmonic assays.

The periodicity of the Au-NPA remains a constant as expected and corresponds to that of the original self-assembled polymeric template. The approach thus allows orthogonal control over the Au_NPA geometry with independent knobs from molecular self-assembly and thin film deposition processes. The strong correlation between Au-NPA structure and the underlying process along with the low standard deviations of the Au-NPA geometry supports the objective of rational design of the plasmonic nanoarrays.

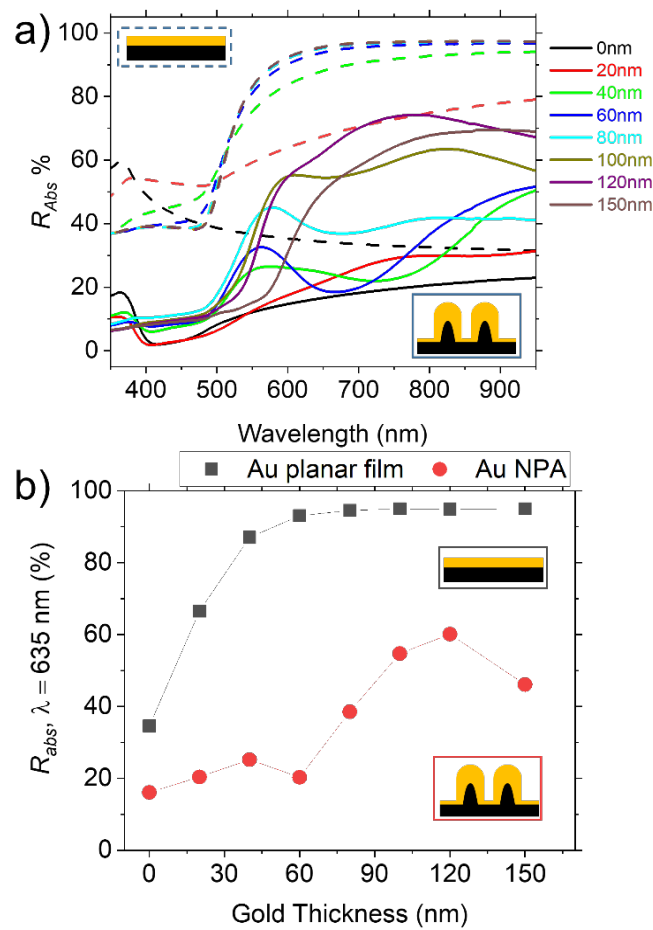

**Figure 3.** Comparison of (a) absolute reflectance spectra of the Au-NPA (continuous line) versus planar gold films (dotted curves) on Si substrate, as a function of gold film thickness, and b) plot of absolute reflectance at 635 nm compared for Au-NPA versus bare substrates.

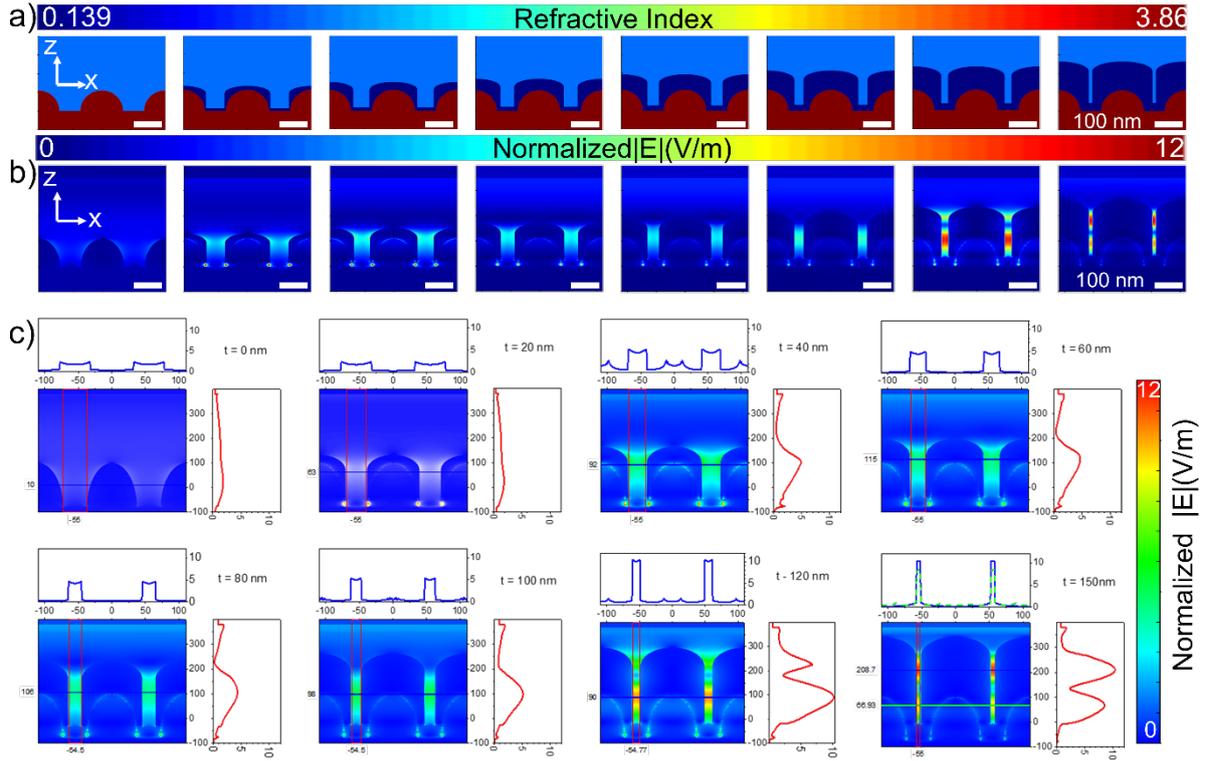

**Figure 4.** (a) Refractive index profile in the x-z cross-section of nanopillar structure. (b) Electric field profile in the x-z cross-section of nanopillar structure at an incident wavelength of 633 nm, and (c) Evolution of electric field profile with increasing gold layer thickness at an incident wavelength of 633 nm. The scale bars indicate 100 nm.

**Inter-Pillar EM Hot-Spots.** The Au-NPA geometry presents a two-fold advantage that make them attractive for plasmonic sensing. First, the nanopillar arrays consist of subwavelength structures that provides an effective refractive-index gradient between air and the substrate, which reduces the reflection of incident light from the nanopillar surface and improves the transmission of light at the interface. This is seen to manifest in the reflectance spectra of the Au NPA which show distinctly reduced reflection of incident light as compared to planar gold film counter parts for all thicknesses. (Figure 3) Anti-reflection and light trapping by textured silicon surfaces has been reported with several profiles in literature. The approach to obtain

Au_NPA in the work has advantage due to opportunity for engineering the optical properties of the interface through rational design of the nanopillar geometries.

Second, the Au-NPA provide high electromagnetic field enhancement at the inter-pillar gaps due to the coupling of localized surface plasmon resonances. This is confirmed by the electric field distribution around nanopillar arrays for different thicknesses obtained by three-dimensional (3D) Finite-Difference Time-Domain (FDTD) simulations. The structural parameters of the nanopillar arrays chosen for the models were guided by the experimental data from Figure 2. The geometric models represent the experimental data closely, and is well supported by the low standard deviation in the different geometric parameters of Au_NPA. (Figure S1, Table S1) In the FDTD algorithm, the anisotropic perfectly matched layer as the absorbing boundary condition is set at z boundary and periodic boundary conditions are set at x and y boundaries of the simulation domain. The grid sizes of 0.5 nm in the x- and z-directions and 1 nm in the y-direction are chosen to ensure numerical convergence. Simulated electric field profiles in Figure 4(bc) confirm expectations of EM hot-spots at the inter-pillar gaps, with increasing enhancements as a function of decreasing pillar separations. The electric field enhancements were found to be maximized at 150 nm gold film thickness, corresponding to an average separation of 6.5 nm for inter-pillar EM hot-spots.

**Molecular Size Dependence.** The sensitivity of Au-NPA towards detection of molecular analytes by SERS was investigated for the role of molecular size in relation to the gap distance of the inter-pillar EM hot-spots. Two different molecular analytes were chosen with molecular size that were an order of magnitude apart, viz. 1-Naphthalene thiol (NT), a small organic molecule with size of ~0.5 nm, and Streptavidin, a tetrameric protein with a molecular weight of 55 kDa, and dimension of 4.2 nm x 4.2 nm x 5.8 nm.[36] SERS spectra were recorded at 633 nm excitation using 50X objective with a numerical aperture of 0.55. NT was used as the small molecular probe, especially due to the well defined spectroscopic characteristics, sub-nanometric size, and its readiness to form self-assembled monolayers on gold surface. The signal intensity of the ring stretching mode of NT at 1377 cm$^{-1}$ was followed to quantify the SERS enhancements as a function of the Au-NPA geometry, and molecular concentrations. Streptavidin is chosen as a model for larger analyte due to well-studied, high-affinity interaction with biotin, with dissociation constant of 10$^{-15}$ M.[37] To detect Streptavidin, the Au-NPA surface was immobilized with biotin-PEG-thiol (BPT), with biotin head groups to bind to Streptavidin (SA) in the solution. The Streptavidin conjugated to Cy5 dye (SA-Cy5) was used for the study, to enable the use of Cy5 as a Raman and Fluorescence reporter to follow the impact of Au-NPA geometry and molecular concentrations on the surface. The SA-Cy5 has an average of 10 Cy5 molecules conjugated to each SA molecule. The SERS intensities of the characteristic Raman peak of Cy5 at 583 cm$^{-1}$ were followed to ascertain the impact of the Au-NPA geometry and the surface concentrations. The SA-Cy5 assays were performed starting with biotinylation of the surface using biotin-PEG thiol, followed by high concentration of BSA to block non-specific binding sites and finally with an exposure to SA-Cy5 of different concentrations.

The impact of Au-NPA geometries, especially the inter-pillar gap EM hot-spots on the SERS intensity evolution of NT and SA-Cy5 was investigated at saturated surface densities of NT and SA-Cy5 obtained respectively at 1 μM solution of ethanol and 400 nM solution in PBS. Under these conditions, the molecular concentrations are large enough in solution to assure complete coverage on surfaces. The use of saturated surface density is essential to ensure that the changes to the SERS signal intensities can be entirely attributed to the Au-NPA geometry. As confirmed by numerical simulations, the Au-NPA present EM gap hot-spots at the interpillar gaps with distances that vary from 30.6 nm – 8.4 nm. Given the molecular dimensions of NT ($d_{NT} \sim 0.5$ nm), and BPT – SA-Cy5 ($d_{BPT} + d_{SA-Cy5} \sim 7.7$ nm), it is reasonable to expect that while NT could access the inter-pillar EM hotspots in all cases, the SA-Cy5 would be unable to access the EM hot-spots with gaps smaller than the interaction dimensions.

The SERS intensity response of NT and SA-Cy5 as a function of the inter-pillar separations is shown in Figure 5.

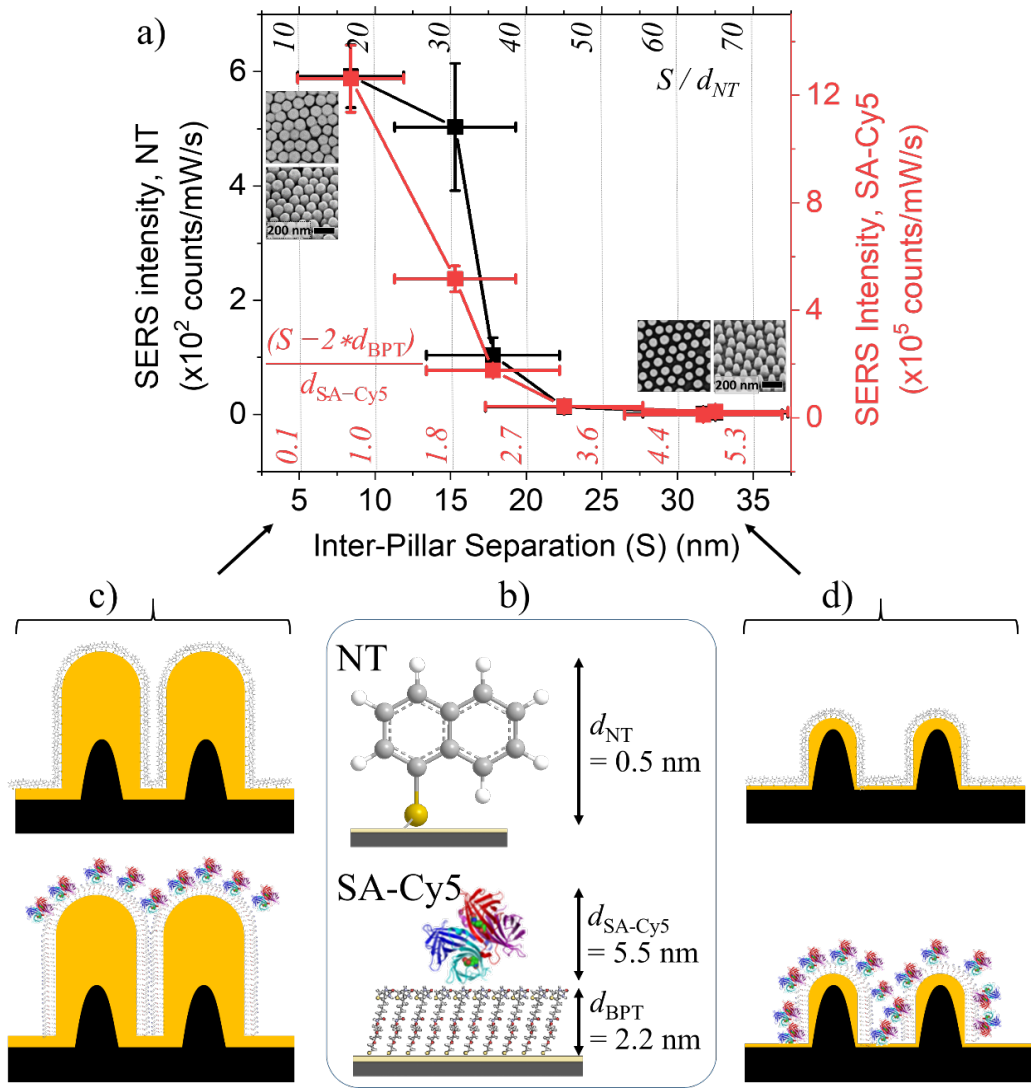

**Figure 5.** Comparison of SERS signal dependence on the inter-pillar separation in Au-NPA for the NT (black curve) and SA-Cy5 (red curve). The inter-pillar gaps are indicated in the units of NT($d_{NT}$) and SA-Cy5 ($d_{SA-Cy5}$) interaction dimensions. The SEM images in top and tilt views of Au-NPA at lowest and highest separations are shown for reference. (b) schematic illustration of the relative dimensions of NT and (SA-Cy5)-BPT pair (c,d) and their expected distribution on Au-NPA at (c) low and (d) high separations.

The SERS intensities for both molecules were matched at the minimum and maximum separations in order to highlight the evolution in SERS response as a function of the gap distance at the EM hot-spots. The inter-pillar separations were further expressed in the units of respective molecular dimensions, thus, the experimental Au-NPA separations could be represented as 4.9 – 0.5 units of SA-Cy5, or 65.2 – 14 units of NT. (Figure 5a) The interpillar gaps in the case of SA-Cy5 detection take into account the reduced gap due to pre-adsorbed self-assembled monolayer of BPT ($d_{BPT} \sim 2.2$ nm), that reduces the separation further by a distance of 4.4 nm.

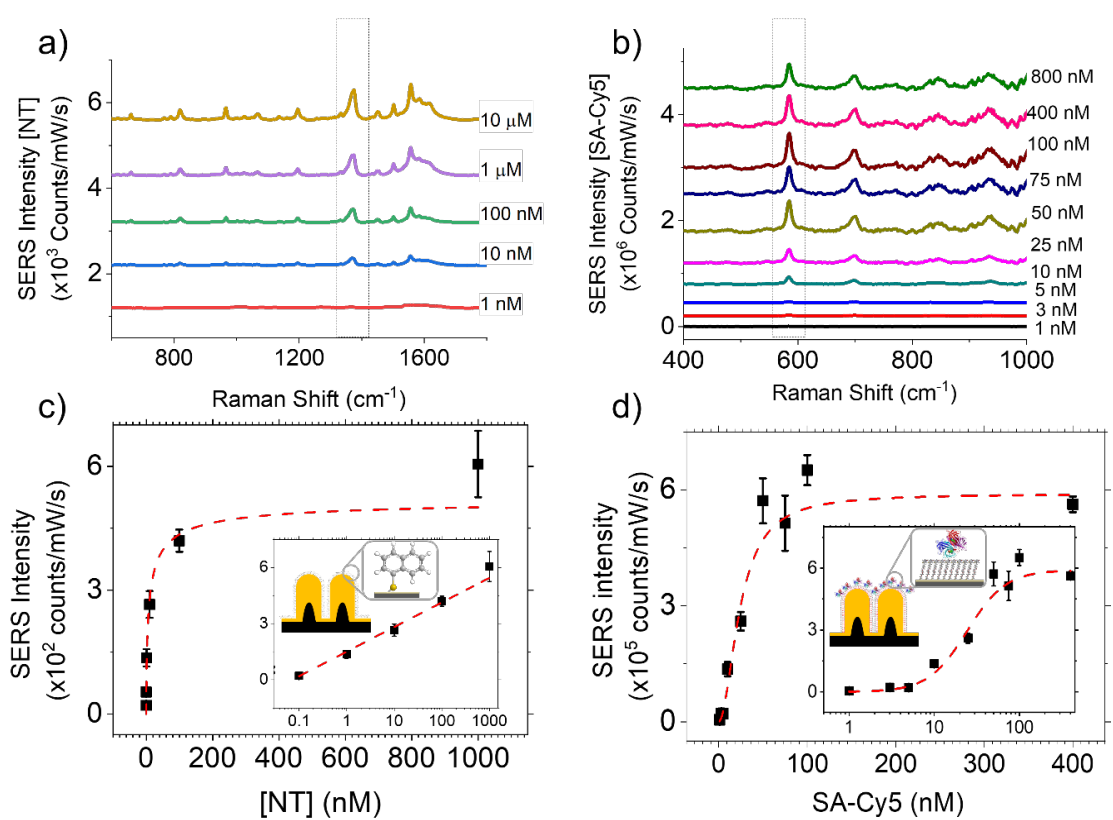

**Figure 6.** Comparison of the SERS assays of (a,c) NT and (b,d) SA-Cy5 on biotinylated Au-NPA_150, showing Raman (a,b) spectral evolution and the (c,d) plot of intensity of characteristic peak of (c) NT (at 1377 cm$^{-1}$) and (d) SA-Cy5 (at 583 cm$^{-1}$) as function of concentration in linear and (inset) logarithmic scales. Dotted lines show Hill-Langmuir fits.

The response as a function of decreasing inter-pillar gaps shows a significant increase in signal intensities below 15.9 nm separations for both NT and SA-Cy5. The rise of the SERS signal intensities to the maximum intensity in the case of NT is steeper compared to that of SA-Cy5. (Figure 5a) The response of the NT continues to increase, but gets less steep between 16.3 nm to 8.4 nm, showing what appears to be signs of saturation. The apparent saturation can be attributed to partial fusion of some of the features at 150 nm thickness as seen in the electron microscopy images, which results from the standard deviation of the pillar periodicities. Despite the low standard deviations of the array periodicities, targeting inter-pillar gaps of the order of the standard deviations would risk closing gaps of a population of pillars corresponding to the lowest separations. The partial fusion reduces the feature densities from 120 $\mu m^{-2}$ to 40 $\mu m^{-2}$ with consequently the density of inter-pillar gaps from ~491 $\mu m^{-2}$ to 161 $\mu m^{-2}$, which is almost 67% reduction in the density of hot-spots. The remaining inter-pillar gaps exhibit lower distances and thus greater EM enhancements, which would explain the continued increase in observed signal intensities.

Unlike in the case of NT, the SA-Cy5 intensities were observed to rise less steeply at pillar separations below 15 nm, which, in other words at gap distances that are below twice the SA-Cy5 dimensions. The gradual increase of SA-Cy5 signal intensities in comparison to that of NT could result from the inability to leverage the inter-pillar EM hot-spots, due to the steric hindrance to entering the gaps. As the pillar gaps reduce further to 8.4 nm, spatial constraints rule out the possibility for any SA-Cy5 at the inter-pillar hot-spots as the gaps can no longer accommodate the molecule. This favors a transition to SA-Cy5 occupying entirely the top of the Au-NPA with the possibility to benefit to the extent of EM enhancements at sites closer to the EM hot-spots. If the arguments were true, the NT would leverage best from its co-localization with inter-pillar EM hot-spots, down to the smallest separations, and should consequently deliver higher analytical sensitivity over SA-Cy5. The SA-Cy5 in comparison would be unable to leverage the best performing EM hot-spots, due to its inability to enter the gap hot-spots, and thus should contribute to lower sensitivity in detection. To confirm this, the Au-NPA with lowest separations (labeled henceforth Au-NPA_150) was evaluated for their performance in quantitative assays of both NT and SA-Cy5. (Figure 6)

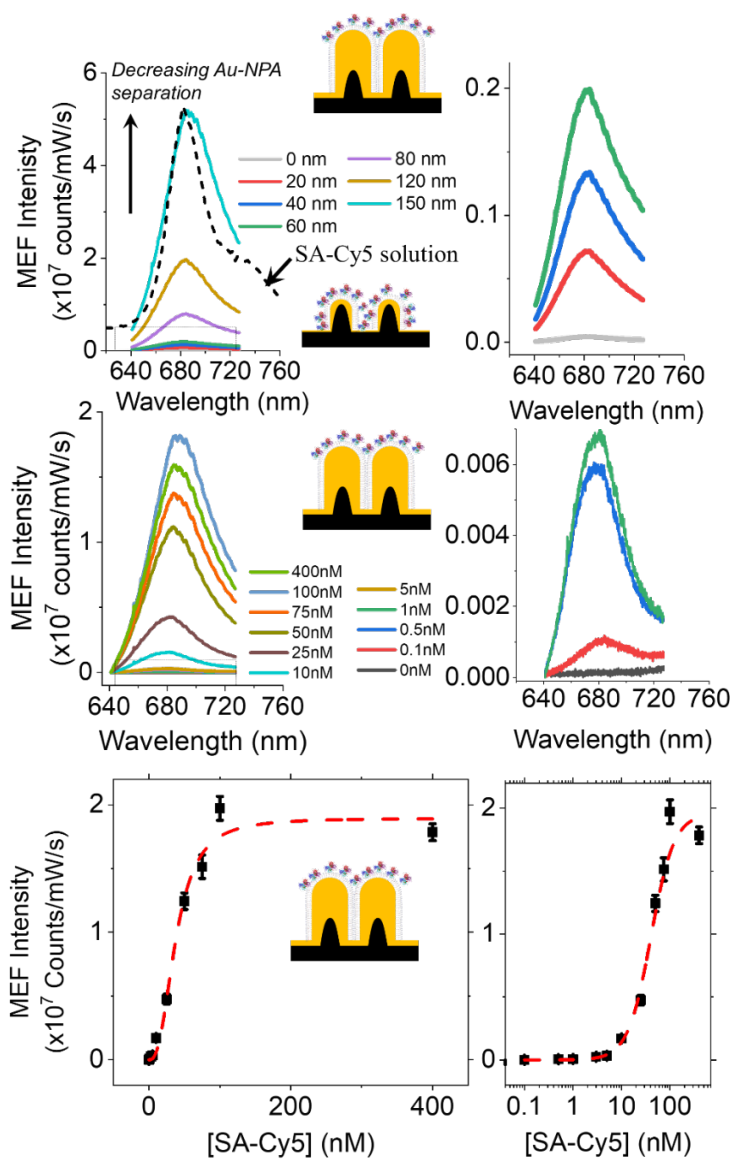

**Figure 7.** a) MEF spectral evolution of saturated coverage of SA-Cy5 on biotinylated Au-NPA as a function of gold thickness. (right) an enlarged representation of the spectra corresponding to lower thickness. b) MEF spectra acquired on Au-NPA_150 as a function of SA-Cy5 concentrations. (right) enlarged representation of spectra corresponding to lower

concentrations. d) MEF assay of SA-Cy5 on biotinylated Au-NPA_150. The dotted line indicates fit to Hill-Langmuir function.

The SERS signal intensities of the characteristic peaks of NT and SA-Cy5 were followed as before as a function of increasing solution concentrations of the respective molecules. The assays show the excellent quantitative response of SERS signal intensities with low standard deviations, even with multiple repetitions of the assays with sensor chips from different batches. The signals for NT on the Au-NPA_150 was observed down to 1nM within the experimental range tested.  The lowest detection limits (LOD) was estimated to be 800 pM taking into account the mean and standard deviation of the blank signal and the standard deviation of the signal intensities at lowest concentration. In case of SA-Cy5, the signals were observed down to 1 nM concentrations, while the LOD was estimated to be 1.3 nM. Considering that each SA-Cy5 has ~10 Cy5 molecules, the LOD corresponds to signal contribution from the Cy5 concentrations that is effectively an order of magnitude higher. The LOD of SA-Cy5 is thus clearly lower as compared to the NT, which is despite the fact that the Cy5 absorption peak (655 nm) is close to that of the excitation wavelength of 633 nm, and thus is capable of resonance Raman excitation. Resonance Raman excitation has the potential to contribute to 3 orders of magnitude higher signal enhancements compared to non-resonant counterparts.[38] The greater sensitivity in the detection of NT is consistent with the observations of the impact of molecular dimensions over gap size-dependent SERS enhancements.

A strategy to overcome the issue of steric limitation to position molecules at the hot-spots is to use metal-enhanced Fluorescence (MEF) in place of SERS. Unlike SERS, MEF requires the analyte to be positioned at an optimal distance away from the hot-spots to benefit from high enhancements.[39–41] This plays to the advantage of partially relieving the spatial constraints associated with SERS. To investigate this, the assays of SA-Cy5 shown in Figure 5 and Figure 6 were repeated while measuring the Fluorescence emission intensities of SA-Cy5 as a function of concentrations. The Fluorescence spectra show strong enhancement in relation to the bare gold substrate controls and increased as a function of gold film thickness with the Au-NPA_150 still exhibiting the maximum signal enhancements amongst the Au-NPAs. (Figure 7a)

The plot of Fluorescence peak intensities of SA-Cy5 gave a well-behaved quantitative response as a function of the concentration of SA-Cy5, with the lowest observable signal at a concentration of 100 pM. The LOD of the assay was estimated to be 120 pM, taking into account the mean and standard deviation of the blank signal and the standard deviation of the signal intensities at 100 pM. (*cf* methods) The LOD observed in the MEF assays is an order of magnitude lower than that of the SERS assays, mainly benefitting from the compatibility of the spatial requirement to the dimension of the molecules. The high sensitivity of the Au-NPA_150 in MEF assays can also be seen from the enhancements by 4 orders of magnitude higher for the Au-NPA_150 as compared to the bare gold surface under identical conditions. The enhancements observed are greater than those reported in the literature, for MEF assays based on Cy5 [41–44] or other dyes,[45,46] and can be attributed to the large number of highly enhancing EM hot-spots attained on the pillar arrays.

**CONCLUSIONS**

Plasmonic nanoarrays consisting of gold nanopillars (Au-NPA) with systematic control over inter-pillar electromagnetic (EM) gap hot-spots were achieved down to sub-10 nm separations. The rational control over the geometric attributes of the NPA, with separations, height, and standard deviations was determined at the early phases of fabrication, due to robust correlations between the process and structure for the different processes used, including self-assembly, pattern-transfer, metal deposition, and surface functionalization. The excellent geometric definition allowed modeling the nanopillar geometries with low deviation from experimental arrays. The numerical simulations show tight confinement of EM fields at the inter-pillar gaps, with maximum enhancements observed for the smallest gap distances. This sets expectations that are consistent with experimental observations of the impact of the molecular size on gap dependent spectral signal evolution. Consequently, the analytical sensitivity in the detection of NT by SERS was an order of magnitude better than the labelled protein analyte, despite the opportunity for resonance Raman excitation in the latter case. The spatial constraints requiring the analyte co-localization at the EM hot-spots for the best leverage in SERS assays could be at least partially relieved by adopting MEF assays. The MEF assays delivered an analytical sensitivity that was an order of magnitude better than the SERS assays pushing the lowest detection limits down to 120 pM. The findings provide convincing case for the need for rational design of nanoplasmonic interfaces to take into consideration, analyte dimensions in relation to EM hot-spots, in order to achieve the highest sensitivities.

**ASSOCIATED CONTENT**

**Supporting Information**

Supporting Information Available: Geometric parameters of the gold nanopillars (Au NP) used for numerical simulations of EM field profiles are tabulated. Plots comparing experimental geometries with those used in the models for numerical simulations are presented. This material is available free of charge via the Internet at http://pubs.acs.org

**AUTHOR INFORMATION**

**Corresponding author**

*E-mail: sivashankar.krishnamoorthy@list.lu

**Author contributions**

The manuscript was written through the contributions of all authors. All authors have given approval to the final version of the manuscript.

**Conflict of Interest**

Authors have no conflict of interest

## Acknowledgments

Funding from the National Research Fund of Luxembourg (FNR) via the project PLASENS (C15/MS/10459961) and partial funding from INTERREG Grand Region V (IMPROVE-STEM) is gratefully acknowledged. SK and RR thank networking and travel funding from EU COST Action BIONECA.

## ABBREVIATIONS

EM, Electromagnetic; NPA, Si Nanopillar Arrays; Au-NPA, Gold coated Nanopillars; SERS, Surface-Enhanced Raman Spectroscopy; MEF, Metal-Enhanced Fluorescence; XPS, X-Ray Photon Spectroscopy; FDTD, Finite-Difference Time-Domain; SEM, Scanning Electron Microscopy; AFM, Atomic Force Microscopy

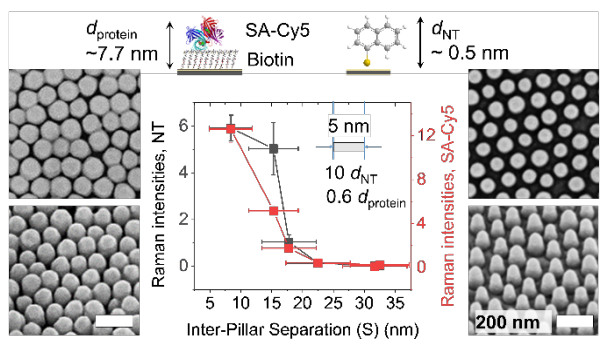